\documentclass[]{aa} 

\usepackage{graphicx}
\usepackage{txfonts}
\usepackage{xcolor}
\usepackage[normalem]{ulem}

\usepackage[breaklinks, colorlinks, citecolor=blue, linkcolor=blue]{hyperref}
\usepackage{listings}

\definecolor{codegreen}{rgb}{0,0.6,0}
\definecolor{codegray}{rgb}{0.5,0.5,0.5}
\definecolor{codepurple}{rgb}{0.58,0,0.82}
\definecolor{backcolour}{rgb}{0.95,0.95,0.92}

\lstdefinestyle{mystyle}{
    backgroundcolor=\color{backcolour},   
    commentstyle=\color{codegreen},
    keywordstyle=\color{magenta},
    numberstyle=\tiny\color{codegray},
    stringstyle=\color{codepurple},
    basicstyle=\ttfamily\scriptsize,
    breakatwhitespace=false,         
    breaklines=true,                 
    captionpos=b,                    
    keepspaces=true,                 
    showspaces=false,                
    showstringspaces=false,
    showtabs=false,                  
    tabsize=2
}

\graphicspath{{/home/aasensio/Dropbox/GIT/DeepLearning/spatially\_variant/validate/figs}}

\newcommand{\vect}[1]{\boldsymbol{\mathbf{#1}}}

\newcommand{\code}{\texttt{SENTIS}}
\newcommand{\codemeaning}{\texttt{SE}mantic \texttt{N}eural \texttt{T}ranslation for the \texttt{I}nversion of \texttt{S}tokes profiles}

\begin{document} 

    \title{Neural translation for Stokes inversion and synthesis}
    
    \author{A. Asensio Ramos\inst{1,2} \and J. de la Cruz Rodr\'{\i}guez\inst{3}}
    
    \institute{
    Instituto de Astrof\'isica de Canarias (IAC), Avda V\'ia L\'actea S/N,
    38200 La Laguna, Tenerife, Spain\\
    \email{andres.asensio@iac.es}
    \and
    Departamento de Astrof\'isica, Universidad de La Laguna, 38205 La Laguna, Tenerife, Spain    
    \and    
    Institute for Solar Physics, Dept. of Astronomy, Stockholm University, AlbaNova University Centre, SE-10691 Stockholm, Sweden
    }

    \date{Received ; accepted }

 
    \abstract
    {The physical conditions in stellar atmospheres, such as temperature, velocity, and magnetic fields, 
    can be obtained from the interpretation of solar spectro-polarimetric observations. However, 
    traditional inversion codes, while successful, are computationally demanding, especially 
    for lines whose formation is complex and dictated by non-local thermodynamical equilibrium effects. The 
    necessity of faster alternatives, particularly with the the increasing volume of data coming from 
    large solar telescopes, has motivated the emergence of machine learning solutions.}
    {This paper introduces an 
    approach to the inversion and synthesis of Stokes profiles inspired by neural machine translation. Our 
    aim is to develop a generative model that treats Stokes profiles and atmospheric 
    models as two distinct ``languages'' encoding the same physical reality. We build a model that learns
    how to translate between them, also providing estimates of the uncertainty.}
    {We employ a tokenization strategy for both Stokes parameters and model atmospheres, which is 
    learned using Vector Quantized Variational Autoencoders (VQ-VAE), a neural model
    used to compress the data into a lower dimensionality form. The core of our inversion code 
    utilizes a transformer encoder-decoder architecture, akin to those used in natural language 
    processing, to perform the translation between these tokenized representations. The model is trained 
    on a comprehensive database of synthetic Stokes profiles derived from perturbations to various semi-empirical solar 
    atmospheric models, ensuring a wide range of expected solar physical conditions.}
    {The method effectively reconstructs atmospheric models from observed Stokes profiles, showing 
    better constrained models within the region of sensitivity of the considered spectral lines. The latent 
    representation induced by the VQ-VAE helps accelerate the inversion by compressing the length of 
    the Stokes profiles and model atmospheres. Additionally, it helps regularize the solution by 
    reducing the chances of obtaining unphysical models. As a final and crucial advantage, 
    the method described in this paper provides the generative nature of our model, which naturally yields an estimate of the uncertainty in the solution.}
    {}
    
    \maketitle
    
    \abstract

%
\section{Introduction}
\label{sec:introduction}
The interpretation of solar spectro-polarimetric observations has given us access
to relevant information about the physical conditions in the solar atmosphere. Information from the temperature
and velocity stratification of the solar atmosphere can be obtained from the analysis of the
intensity of selected spectral lines. Information about the magnetic field vector can be grasped from the
analysis of the polarization profiles of these lines. In the solar spectrum we find relatively weak lines 
that have a formation region in the photospheric regions, and their interpretation gives us information about photospheric
temperatures, velocities and magnetic fields. Stronger lines, that are formed in the chromosphere and
transition region, provide us with information higher up in the solar atmosphere.

The interpretation of the observations is usually done by means of inversion codes \citep[see][]{2016LRSP...13....4D,2017SSRv..210..109D}. 
These codes solve the radiative transfer equation in a predefined parametric model of the solar atmosphere and
iteratively adjust the parameters of the model to match the observed profiles. The success of such
inversion codes is doubtless given the enormous results produced in the last few decades. The simplest codes are based
on the Milne-Eddington atmosphere, which provides an analytical solution to the radiative transfer equation
\citep{1977SoPh...55...47A,2004ASSL..307.....L}. Examples of these codes are 
\texttt{MERLIN} \citep{2007MmSAI..78..148L}, \texttt{MILOS} \citep{2007A&A...462.1137O}, 
\texttt{VFISV} \citep{2011SoPh..273..267B}, and \texttt{pyMilne} \citep{2019A&A...631A.153D}.
More complex models require the prescription of gradients along the line-of-sight (LOS). The first code
based on this idea was the Stokes Inversion based on Response functions code \citep[\texttt{SIR};][]{1992ApJ...398..375R}, 
which demonstrated for the first time that it is possible to infer full stratifications of the temperature, velocity 
and magnetic field vector via the inversion of the Stokes profiles. \texttt{SIR} was limited to lines whose formation 
is modeled under the local thermodynamic equilibrium (LTE) approximation, which limits it to relatively
weak photospheric lines. A handful of codes have been developed since following the same strategy, among which
some of the most successful are \texttt{SPINOR} \citep{2000A&A...358.1109F}, \texttt{FIRTEZ} \citep{2019A&A...629A..24P}, 
\texttt{SPINOR-2D} \citep{2012A&A...548A...5V} and \texttt{Hazel2} \citep{2008ApJ...683..542A}, the latter 
simply building a wrapper around SIR.
The inversion of stronger chromospheric lines requires the use of non-LTE codes \citep{2000ApJ...530..977S}
which need to iteratively solve 
the radiative transfer equation in consistency with the statistical equilibrium equations at each iteration of 
the inversion process. For this reason, these codes are much more complex and computationally demanding.
Examples of these codes are \texttt{NICOLE} \citep{2015A&A...577A...7S}, \texttt{STiC} \citep{2019A&A...623A..74D},
\texttt{SNAPI} \citep{2018A&A...617A..24M} and \texttt{DeSIRe} \citep{2022A&A...660A..37R}.

The availability of large solar telescopes like the Swedish Solar 
Telescope \citep[SST;][]{2003SPIE.4853..341S,2024A&A...685A..32S}, the GREGOR 
telescope \citep{GREGOR, kleint2020}, Hinode SOT/SP \citep{lites_hinode01,kosugi_hinode07},
or the Daniel K. Inouye Solar Telescope \citep[DKIST;][]{2020SoPh..295..172R} have provided us with
enormous amounts of data. If the inversion of a single pixel can take several seconds in LTE or several
minutes in non-LTE, the inversion of a full field-of-view (FoV) with tens of thousands of pixels
can take several hours or days, even when paralellizing the inversion process in supercomputers. 
This had led, thanks to the revolution of modern machine learning (ML) techniques, to the development of
ML-based inversion codes that can provide results in a fraction of the time required by traditional
approach, also providing additional benefits. The first such approach was proposed by \cite{2001A&A...378..316C},
who proposed the use of multi-layer perceptrons (MLP) for the inversion of Stokes profiles under
the Milne-Eddington approximation. Once trained with synthetic Stokes profiles,
the neural network can provide a solution to the inverse problem very fast. \cite{2003NN.....16..355S}
took advantage of this strategy to measure magnetic fields in sunspots while 
\cite{2005ApJ...621..545S} proposed strategies to improve the results of the neural inversion.
A step forward was carried out by \cite{2008A&A...481L..37C}, who demonstrated that neural networks
can also be used to quickly infer the depth stratification of the physical conditions from lines formed
in LTE.

A significant advance in the field was achieved by \cite{2019A&A...626A.102A}, who built the Stokes Inversion based
on COnvolutional Neural networks (\texttt{SICON}) code. \texttt{SICON} uses convolutional neural networks (CNN) to
infer the mapping between the Stokes parameters and the stratification of the physical conditions in the solar 
atmosphere for Hinode SOT/SP observations. \texttt{SICON} was trained with a large database 
of photospheric synthetic Stokes profiles generated from numerical
simulations of the solar atmosphere and can provide results in a fraction of a second for a large FoV, 
many orders of magnitude faster than traditional inversion codes. CNNs take advantage of
the spatial and spectral correlation in the observations to provide clean maps of physical quantities. They can
additionally exploit statistical correlations in the training set to provide an estimation of the physical
conditions in geometrical height, also decontaminating the observations from the blurring effect of the
telescope.

The revolution provided by the advent of ML seems unstoppable in science and, in particular, in 
the field of spectropolarimetric inversion of Stokes profiles \cite[see][]{2023LRSP...20....4A}. 
\cite{2024ApJ...976..204Y} have recently
built \texttt{SPIn4D}, a CNN approach similar to \texttt{SICON} to carry out fast inversions 
for spectropolarimetric data observed with DKIST. \texttt{SuperSynthIA} \citep{2024ApJ...970..168W} produces
full-disk vector magnetograms from Stokes vector observations of the Helioseismic and Magnetic Imager 
(HMI) on board the Solar Dynamics Observatory \citep[SDO;][]{2012SoPh..275..207S}. One-dimensional 
CNNs have also been used to accelerate the inversion of Stokes profiles \citep{2020A&A...644A.129M,2022ApJ...940..147L}
and also to provide better initializations for classical inversion codes \citep{2021A&A...651A..31G}. MLPs 
are also still very useful, as demonstrated by \cite{2021A&A...652A..78S}, who used neural inversions to 
provide direct observational evidence of the hot wall effect in magnetic pores.
As a path towards explainability, \cite{2022ApJ...925..176C} investigated the physical content learned by CNN
when trained to carry out the inversion of Stokes profiles, demonstrating that CNNs are able to 
exploit the physically meaningful relation between wavelengths and geometrical heights in the solar atmosphere.
Recently, \cite{arxiv2506.16810} has explored the use of transformers \citep{NIPS2017_3f5ee243} to carry out 
the inversion of Stokes profiles, showing that they can outperform MLPs.

Instead of using neural networks to simply map the inverse problem directly, other more 
elaborate approaches have also been pursued.
One example is the use of neural fields (using neural networks to map coordinates on the space 
or space-time to coordinate-dependent field quantities) as flexible 3D models of the solar atmosphere, which
are then learned from observations using the known physics of spectral line formation. 
\cite{2025A&A...693A.170D} successfully used the weak-field 
approximation model to simulate the Stokes profiles observed at all points in a FoV parameterized by the neural field,
which is then trained by minimizing the difference between the observed and synthetic Stokes profiles. Later,
\cite{2025arXiv250213924J} increased the complexity by assuming a Milne-Eddington model, also showing
promising results. Efforts
are also being made to use neural fields to model the full 3D solar atmosphere and infer them using
lines in LTE or non-LTE. The fact that neural fields need to be trained by backpropagating gradients
through the radiative transfer solution makes them computationally expensive, especially in non-LTE 
conditions. More effort needs to be put into investigating whether adjoint methods 
can accelerate the training. Emulators that speed up the radiative transfer solution are also being in exploration as 
a solution to this problem.

A crucial aspect of the inversion of solar spectro-polarimetric observations, many times overlooked, 
is the estimation of the uncertainty in the solution. The inversion of the Stokes profiles is, in general,
an ill-posed problem with non-unique and ambiguous solutions. Although inversion codes can provide a point estimate of the 
covariance matrix at the solution \citep[e.g., see][]{1992ApJ...398..375R}, the uncertainties
obtained from this covariance matrix tend to be unreliable because they are calculated purely based 
on the goodness of the fit and the parameter sensitivity of the problem. One has to rely then on time consuming Bayesian approaches which provide
a full posterior distribution over the parameters. \cite{2007A&A...476..959A} was the first study to
introduce a Bayesian approach to the inversion of Stokes profiles using Markov Chain Monte Carlo (MCMC) methods.
The implementation of ML solutions is allowing us to accelerate this process by a large margin, sometimes
orders of magnitude. \cite{2019ApJ...873..128O} used an invertible neural network \citep{2018arXiv180804730A} 
to capture degeneracies and uncertainties
in the inference of the physical conditions in flaring regions. \cite{2022A&A...659A.165D} used
normalizing flows to provide very fast variational approximations to the posterior distribution in LTE and non-LTE
inversions. \cite{2023SoPh..298...98M} used a different approach in which a CNN is trained to
provide a point estimate of the mean and standard deviation of the physical parameters. The neural
network is trained via maximum-likelihood, assuming a Gaussian likelihood. Finally, \cite{2024ApJ...977..101X}
used an MLP as an emulator to sample from the posterior distribution of the physical parameters
using Hamiltonian Monte Carlo (HMC) methods.

\begin{figure*}
  \centering
  \sidecaption
  \includegraphics[width=\textwidth]{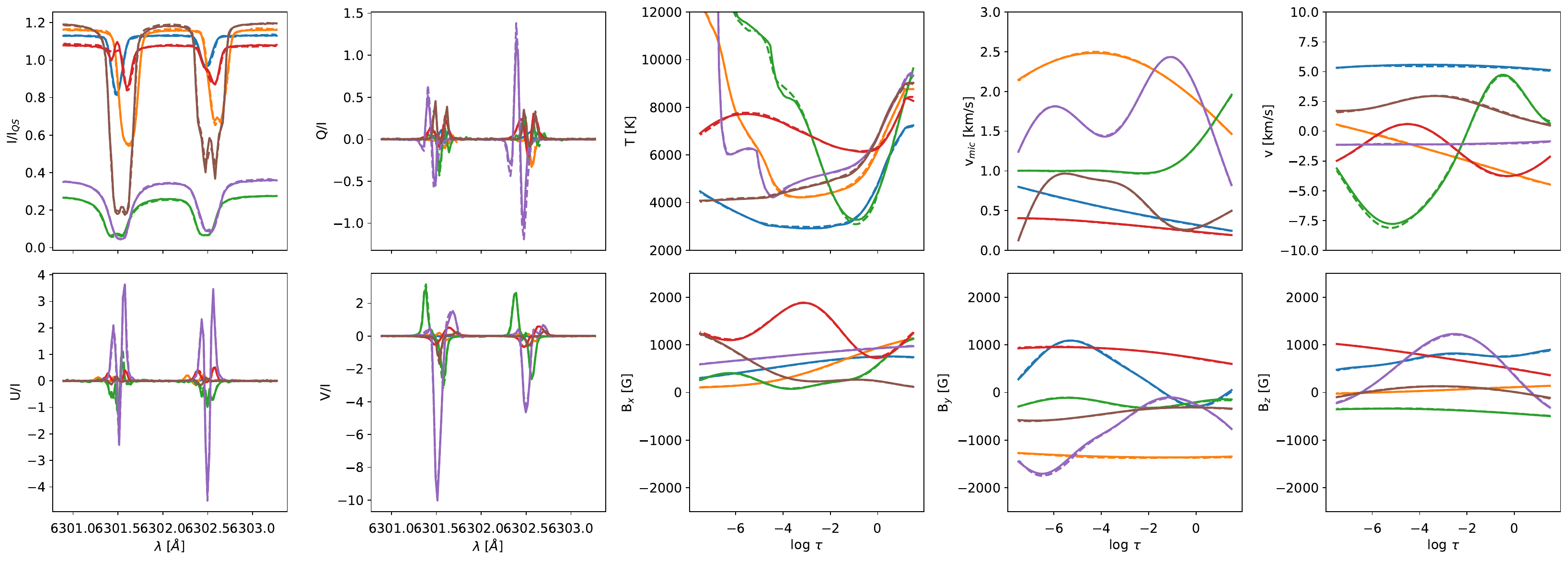}
  \caption{Examples of the Stokes profiles (first two columns) and the model atmospheres (last three columns)
  used for training \code. Solid lines represent the original data, while the dashed lines represent the
  reconstructed ones using the VQ-VAE model.}
  \label{fig:training_set}
\end{figure*}

On a different, apparently unrelated note, the field of natural language processing (NLP) 
has seen a revolution in the last few years by the
introduction of autoregressive generative models. These models predict one word, syllable, or token in general, from
the previous ones given some context information. Inspired by the success of these models in NLP, 
we present in this work a new approach to the 
inversion of Stokes profiles based on the ideas of neural machine translation. For the sake of shortness, in the following we
refer to the method as \code\ (from \codemeaning)\footnote{\texttt{https://github.com/aasensio/sentis}}.
This requires a new perspective in which Stokes profiles and model atmospheres are considered 
as two different ``languages'' that can be used to represent the same 
underlying physical reality. To this end, we require to establish a dictionary for each language using 
a tokenization approach. Under these assumptions, we can train a generative model that learns how to translate 
one language into the other, and viceversa. Autoregressive models trained to predict the next token in a 
sequence are stochastic by nature. As a consequence, they are generative models that can 
be used to provide a measure of uncertainty in the 
translation. One key aspect of this tokenization approach is that it does not require a parametric model of the solar atmosphere.
Both models and Stokes profiles are represented in tokenized forms, which allows us to infer 
a large variety of model atmospheres.

\section{Training data}
\label{sec:training}

It is a well-known fact that deep learning (DL) models improve their performance with the amount of training data.
The performance of generative models have been shown to improve with the amount of training data, the number 
of parameters in the model and the amount of compute used to train it, as shown by the 
so-called scaling laws \citep{2020arXiv200108361K,2020arXiv201014701H}.
For this reason, the results shown in this paper have been obtained with a large database of synthetic 
Stokes profiles and model atmospheres. The model atmospheres are defined by the stratification
of the temperature ($T$), the line-of-sight (LOS) velocity ($v$), the microturbulent velocity ($v_\mathrm{mic}$),
and the three cartesian components of the magnetic field vector ($B_x$, $B_y$, $B_z$). We consider the
physical parameters sampled at a common grid of 80 equispaced points in the logarithm of the optical depth 
at 500 nm ($\log \tau_{500}$), from $-7.5$ to $1.5$. The emergent Stokes profiles are computed in the
pair of Fe \textsc{i} lines at 630 nm as representative of a photospheric line and in the Ca \textsc{ii}
infrared line at 854 nm, as representative of a chromospheric line. The Stokes profiles of the 
Fe \textsc{i} lines are computed from the model atmospheres in 112 wavelength points
between 630.08921 and 630.32671 nm. This is the range covered by the Hinode SOT/SP instrument. We 
use \textsc{Hazel2} for the synthesis, under the assumption of LTE. Concerning the Ca \textsc{ii} line, we compute 
the Stokes profiles in 96 wavelength points between 854.108 and 854.309 nm. To compute the 
Stokes profiles we use \texttt{STiC}, which takes non-LTE effects into account through its 
\texttt{RH} synthesis module \citep{2001ApJ...557..389U}.

\begin{table}[t]
  \centering
  \caption{Considered base semi-empirical models.}
  \label{tab:models}
  \begin{tabular}{lcc} 
    \hline\noalign{\smallskip}
  Name /  & Region & Reference\\ 
    \hline\noalign{\smallskip}
  HOLMU & Quiet Sun & \cite{1974SoPh...39...19H}\\
  HSRA & Quiet Sun & \cite{1971SoPh...18..347G}\\
  VAL-C & Quiet Sun & \cite{1981ApJS...45..635V}\\
  MACKKL & Sunspot & \cite{1986ApJ...306..284M}\\
  grevesse & Quiet Sun & \cite{1999A+A...347..348G}\\
  emaltby & Sunspot & \cite{1986ApJ...306..284M}\\
  mmaltby & Sunspot & \cite{1986ApJ...306..284M}\\
  cool & Sunspot & \cite{1994A+A...291..622C}\\
  hot & Sunspot & \cite{1994A+A...291..622C}\\
  penumjti & Sunspot & \cite{1994ApJ...436..400D}\\
  solannt & Network & \cite{1986A+A...168..311S}\\
  solanpl & Plage & \cite{1992A+A...263..312S}\\
  FAL-B & Quiet Sun & \cite{2009ApJ...707..482F}\\
  FAL-C & Quiet Sun & \cite{2006ApJ...639..441F}\\
  FAL-D & Network & \cite{2009ApJ...707..482F}\\
  FAL-E & Network & \cite{2006ApJ...639..441F}\\
  FAL-F & Network & \cite{2009ApJ...707..482F}\\
  FAL-H & Plage & \cite{2009ApJ...707..482F}\\
  FAL-P & Facula & \cite{2009ApJ...707..482F}\\
  FAL-R & Penumbra & \cite{2009ApJ...707..482F}\\
    \hline
  \end{tabular}
\end{table} 

\subsection{Generation of training data}
\label{sec:training_data}
In order to generate the training data, we consider the semi-empirical models of the solar atmosphere that are
listed in Table~\ref{tab:models}, from which we only keep the temperature. These temperature
stratifications have been obtained from the interpretation of different observations (see the references in the table for
details) and are widely used in the literature. We use them because they cover a wide range of physical conditions, 
from the quiet Sun to sunspots. They are considered as base models over which we randomly add
perturbations, similar to what is done in \cite{2021A&A...652A..78S}. For this 
exploratory work, we have only considered perturbations described by a Gaussian process \citep[GP;][]{Rasmussen2006Gaussian} 
with a squared exponential covariance kernel, which generate perturbations that vary relatively smoothly in optical
depth. We will consider other options in the future, from using stratitications extracted from magneto-hydrodynamical 
simulations to non-continuous stratifications that can be used to represent shocks. 

The perturbations $\vect{\Delta}$ at all depths in the discretized 
atmosphere are given by the zero mean multivariate Gaussian distribution $p(\vect{\Delta})=\mathrm{N}(0,\vect{\Sigma})$, where
the matrix elements of the covariance matrix are given by the following function\footnote{The diagonal term is used to
stabilize the inversion of the covariance matrix, which is otherwise ill-conditioned.}:
\begin{equation}
  \Sigma_{ij} = \sigma_i \sigma_j \exp\left(-\frac{(\log \tau_{500,i}-\log \tau_{500,j})^2}{l^2}\right) + 10^{-8} \sigma_i \sigma_j \delta_{ij}.
\end{equation}
Here, $\sigma_i$ is the standard deviation of the perturbation at depth $i$, and $l$ is the length scale of the perturbation.
In our case, we find good results using $\sigma_i(T)$ that linearly goes from 3000 K at $\log \tau_{500}=-7.5$ to 1000 K 
at $\log \tau_{500}=1.5$, and $\sigma_i(v)=3$ km s$^{-1}$, $\sigma_i(v_\mathrm{mic})=1$ km s$^{-1}$, 
and $\sigma_i(B_x)=\sigma_i(B_y)=\sigma_i(B_z)=700$ G, all of them constant throughout the atmosphere. In order to avoid perturbations that are too far
from those expected in the solar atmosphere, we discard samples that are not in the range $[2500, 150000]$ K for 
the temperature, $[-30, 30]$ km s$^{-1}$ for the LOS velocity,
$[0, 5]$ km s$^{-1}$ for the microturbulent velocity, $[-3500, 3500]$ G for the $B_y$ and $B_z$ components of 
the magnetic field, and $[0, 3500]$ G for the $B_x$ component of the magnetic field. This last condition is imposed
to remove the 180-degree ambiguity in the azimuth of the magnetic field vector.
Concerning the length scale $l$, each model in the training set is obtained using a different value, extracted
uniformly in log scale in the range $[2, 25]$. Models with large $l$ are very smooth, while models with small $l$ have
structure at much smaller scales. After the synthesis, we found a fraction of the models that produce 
values in the continuum that are too high with respect to what is observed in the Sun. We discard models
whose continuum intensity at 630 nm is above 1.3. Although we generate 1 million models, only 0.61 million passes this
condition and are used for training. An additional set of 6000 models are used for validation and another set of 6000 for testing.
Examples of the models are shown in the third, fourth and fifth columns 
of Fig.~\ref{fig:vqvae}. The emergent Stokes profiles for the Fe \textsc{i} doublet are displayed in the first two columns of the figure.

\section{Autoregressive neural machine translation}
\label{sec:architecture}
We consider the inference of atmospheric parameters from the interpretation of
the Stokes profiles as a machine translation problem. Given a sequence $\mathbf{S}=\{S_1,S_2,\ldots,S_N\}$ 
of length $N$ that represents the Stokes parameters (be it the concatenated four Stokes parameters
or a specific tokenization of them),
the purpose of the model is to provide a probability distribution $p(\vect{\theta}|\mathbf{S})$ over 
a sequence $\vect{\theta}$ of length $M$ that represents the model atmosphere (be it the concatenated 
stratifications or a specific tokenization of them). Having access to the 
probability distribution produces a generative model that can be used to sample from the posterior distribution.
It is customary to use autoregressive models to solve this problem, which encode the 
full posterior distribution as a product of conditional distributions:
\begin{equation}
  p(\vect{\theta}|\mathbf{S}) = \prod_{i=1}^M p(\theta_i|\theta_1,\ldots,\theta_{i-1},\mathbf{S}).
\end{equation}
This means that the model is trained to predict the next token in the sequence given 
all previous tokens and the information about the Stokes parameters.
The translation is typically done using an encoder-decoder architecture. An encoder uses $\mathbf{S}$
as input and produces a latent representation that is assumed to encode all the relevant
information needed for the translation. The decoder takes this latent representation and produces,
in an autoregressive way, an estimation of $p(\theta_i|\theta_1,\ldots,\theta_{i-1},\mathbf{S})$.

If the model $\vect{\theta}$ is chosen to be the full depth stratification of the model atmosphere, 
i.e., $\vect{\theta}=\{\mathbf{T},\mathbf{v},\mathbf{v}_\mathrm{mic},\mathbf{B}_x,\mathbf{B}_y,\mathbf{B}_z\}$,
where $\mathbf{x}=\{x(\log \tau_{500,1}),\ldots,x(\log \tau_{500,M})\}$ for all variables $x$,
one faces a regression problem because the model $\vect{\theta}$ is continuous. In such case, one 
needs to pre-define a functional form for the 
probability distribution $p(\theta_i|\theta_1,\ldots,\theta_{i-1},\mathbf{S})$, something that is not
known in advance. For this reason, we consider in this work a different approach. We
use a tokenization of the model atmosphere and the Stokes parameters (although that of the Stokes
parameters is not mandatory). This tokenization introduces two desirable side effects. First, 
it is a very efficient way of compressing the information in the
model atmosphere (and Stokes parameters) using a finite set of features. 
The power of tokenization lies in the fact that this compression is learned, as 
described in Sec. \ref{sec:tokenizers}, from the training data 
and allows us to represent the model atmosphere and the Stokes parameters
as finite sequences of tokens extracted from a fixed and finite codebook. The second side
effect is that the tokenization allows us to transform the problem into a classification one, so that
we can use a categorical distribution for the conditional probabilities:
\begin{equation}  
  p(\theta_i|\theta_1,\ldots,\theta_{i-1},\mathbf{S}) = \mathrm{Cat}(\theta_i|\theta_1,\ldots,\theta_{i-1},\mathbf{S}),
  \label{eq:categorical}
\end{equation}
where $\mathrm{Cat}(\theta_i|\theta_1,\ldots,\theta_{i-1},\mathbf{S})$ is the categorical distribution over 
the set of tokens that represent the model atmosphere. Treating the problem as a categorical or
classification one allows us to easily define the full posterior distribution and 
end up with a properly defined generative model.

We note that the translation problem is completely symmetric, so it is straighforward to also consider the
inverse problem. In such case, we consider the Stokes parameters as the target sequence and the model atmosphere
as the input sequence. We consider this option in Sec. \ref{sec:synthesis}, where we end up
with a generative model that can be used to produce emulated synthetic Stokes profiles from a model atmosphere,
equipped with an uncertainty estimate.

\subsection{Tokenizers}
\label{sec:tokenizers}
The wavelength variation of the Stokes parameters contain lots of 
redundant data and are strongly compressible \citep[see, e.g.,][]{2007ApJ...660.1690A}. Dimensionality 
reduction techniques like principal component analysis (PCA) or autoencoders can be used to compress the
Stokes parameters into a smaller set of features 
\citep{2000A&A...355..759R,2005ApJ...620..517S,2005ApJ...622.1265C}. The same idea applies to
the model atmosphere, as demonstrated by the success of inversion codes like \texttt{SIR}, which parameterize
the model atmosphere with only a few nodes. 

\begin{figure}
  \centering
  \sidecaption
  \includegraphics[width=\columnwidth]{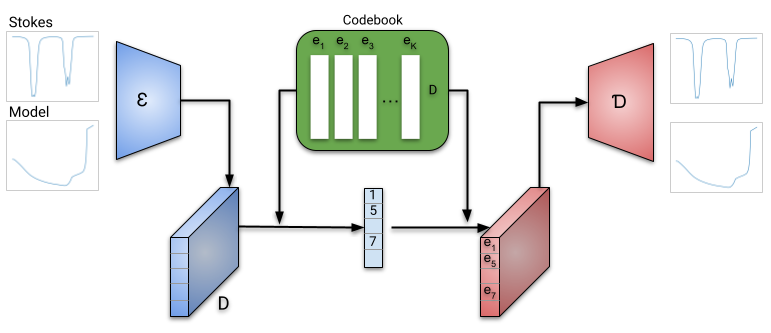}  
  \caption{VQ-VAE model used for tokenization. The encoder $\mathcal{E}$ produces a latent
  representation of the Stokes parameters and the model stratifications, which is then 
  passed through the decoder $\mathcal{D}$ after quantization.}
  \label{fig:vqvae}
\end{figure}

Since classical dimensionality reduction methods like PCA or autoencoders are not suitable 
for the auto-regressive
translation problem because they provide a continuous representation of the data, 
we draw upon VQ-VAE \citep{2017arXiv171100937V}
to learn a tokenization of the Stokes parameters and the model atmosphere. VQ-VAEs are a type of
variational autoencoder (autoencoder with a prescribed prior on the latent space) that learns a 
discrete representation of the data by mapping the continuous latent space produced
by the encoder to a finite set of discrete tokens from a learned codebook (sef Fig. \ref{fig:vqvae}). 
The encoder $\mathcal{E}$ maps the input data (every Stokes parameter and physical variable, properly
normalized) to a latent representation of reduced spatial size 
but increased dimensionality ($D$). The encoder first applies a 1D CNN layer with a kernel of 
size 3 that projects the 1D vector into a space of dimension 64 for the model parameters and 256 for 
the Stokes parameters. We checked that using a lower value for the Stokes parameters did not
produce sufficiently good reconstructions. A consecutive application of three blocks containing a 1D convolutional layer with a kernel of
size 3, GELU activation function \citep{2016arXiv160608415H}, another 1D convolutional layer with a kernel of size 3
and a max pooling operation, is applied to reduce the spatial size of the vectors. A final 1D 
convolutional layer with a kernel of size 1 projects the latent representation to the desired
dimension $D=8$. Since the Stokes parameters for the Fe \textsc{i} spectral region contains 112 wavelength points, 
we end up with vectors of size $112/2^3=14$. Likewise, the Ca \textsc{ii} spectral region contains 96 wavelength
points, so they are reduced to vectors of size $96/2^3=12$. The model parameters are defined at 80 optical depth points, so the spatial dimension is $80/2^3=10$. 
We point out that we have not carried out a systematic search for the optimal
specific arquitectures of the encoder and decoder, since we only aim to demonstrate the feasibility of the approach. 
An architecture based on transformer encoders can potentially deal with Stokes parameters and
model parameters sampled at arbitrary wavelengths and optical depths, but this is left for future work.

\begin{table}[t]
  \centering
  \caption{Normalization of the Stokes parameters and the model parameters before entering the VQ-VAE used
  for tokenization.}
  \label{tab:normalization}
  \begin{tabular}{lcc} 
    \hline\noalign{\smallskip}
  Parameter & $x_\mathrm{min}$ & $x_\mathrm{max}$\\ 
    \hline\noalign{\smallskip}
  $I/I_\mathrm{QS}$ & 0 & 2.5\\
  $Q/I$ & -0.01 & 0.01\\
  $U/I$ & -0.01 & 0.01\\
  $V/I$ & -0.01 & 0.01\\
  $T$ [K] & 2000 & 25000\\
  $v$ [km s$^{-1}$] & -10 & 10\\
  $v_\mathrm{mic}$ [km s$^{-1}$] & 0 & 3\\
  $B_x$ [G] & -1000 & 1000\\
  $B_y$ [G] & -1000 & 1000\\
  $B_z$ [G] & -1000 & 1000\\
    \hline
  \end{tabular}
\end{table} 

The latent representation is then quantized to the nearest token in the codebook $\mathcal{C}$, which is a set of $K=256$
discrete vectors of dimension $D=8$. The quantization is done by computing the $\ell_2$ distance between the latent representation
and all the tokens in the codebook, and selecting the token with the smallest distance. This means that every Stokes 
parameter is described by a sequence of 14 tokens for the Fe \textsc{i} doublet and 12 tokens for
the Ca \textsc{ii} line, extracted from the codebook. The same applies to the model parameters, which
are described by a sequence of 10 tokens. These tokens play a role similar to the nodes used in 
inversion codes like \texttt{SIR}. The advantage is that, once learned, the tokenization can produce stratifications 
with the whole range of complexity and variability present in the training set, without the need to 
prescribe the number of nodes for the model atmosphere. 
As a final step, the quantized representation is then passed to the 
decoder $\mathcal{D}$, which is a mirror of the encoder, using a linear upsampling operation instead of max 
pooling to increase the spatial size of the vectors. Both the Stokes parameters and the model parameters 
are normalized before being encoded with the VQ-VAE with the following formula:
\begin{equation}
\bar{x} = 2 \frac{x-x_\mathrm{min}}{x_\mathrm{max}-x_\mathrm{min}} - 1.
\end{equation}
The values of $x_\mathrm{min}$ and $x_\mathrm{max}$ for each parameter are shown in Tab. \ref{tab:normalization}.

The VQ-VAE is trained to minimize the reconstruction error between the input data and 
the output of the decoder, measured with the mean squared error, while making efficient use of the discrete 
codebook \cite[for details, see][]{2017arXiv171100937V}. Noise with 
a standard deviation of $10^{-3}$ in units of the continuum intensity is added to the 
Stokes profiles during training. No noise is added to the atmospheric models. The VQ-VAE is trained with 
a batch size of 256 using the Adam optimizer \citep{kingma2014adam} with a learning rate of $3 \times 10^{-4}$, which 
is reduced following a 
cosine annealing law until reaching $3 \times 10^{-5}$ after 30 epochs. The VQ-VAE does a very good job in 
reproducing the Stokes parameters and the model parameters from the quantized representation, as shown in
Fig. \ref{fig:training_set}. The dashed curves in Fig. \ref{fig:training_set} show the reconstructions, which 
almost perfectly overlap the original curves.

We note that the tokens of the physical parameters obtained by VQ-VAE are ordered starting 
from the deepest layer in the solar atmosphere, i.e., the first token corresponds to the
quantities at $\log \tau_{500}=1.5$. This is convenient for the autoregressive translation model, as the
variability of the models in the deep layers is much smaller than in the upper layers.

\subsection{Transformer encoder-decoder}
\label{sec:transformer}
The transformer encoder-decoder architecture \citep{NIPS2017_3f5ee243} is the backbone of the autoregressive
translation model. The encoder is a stack of four transformer encoder blocks, each of which
contains a multi-head self-attention layer and a feed-forward neural network. The self-attention layer allows the model
to learn the relationships between the tokens in the input sequence, while the feed-forward neural network
applies a non-linear transformation to the output of the self-attention layer. We have also not done
a systematic search for the optimal hyperparameters of the transformer architecture, but we have
used 8 heads in the self-attention layer, and the feed-forward neural network uses a hidden layer of size 1024.

The input to the encoder is the concatenation of the token indices computed with the VQ-VAE for each one of the four Stokes 
parameters. The total sequence length is then 56 tokens when the Fe \textsc{i} spectral region is used and 48 if only 
the Ca \textsc{ii} spectral region is used. If more than one spectral region is used, we simply concatenate 
the token indices of the different spectral lines. In our case, the maximum sequence length is 104 when we consider the
two spectral regions. Although the
number of input tokens in our case is always fixed in this work, we prefer to use specific tokens to indicate
the start-of-sentence (\texttt{SOS}) and end-of-sentence (\texttt{EOS}), in anticipation of future work that will
require the model to handle variable-length sequences when using arbitrary combinations of different 
spectral lines at arbitrary wavelength samplings. The encoder then uses an internal (learned) 
embedding representation with $256+2$ vectors of dimension $D=256$. The extra two vectors are 
associated with the \texttt{SOS} and \texttt{EOS} tokens.

We add the standard positional encoding \citep[see, e.g.,][]{NIPS2017_3f5ee243}
to the input sequence to provide the model with information about the order of the tokens in the sequence:
\begin{align}
  \mathrm{PE}_{j,2i} &= \sin\left(\frac{j}{10000^{2i/D}}\right), \nonumber \\
  \mathrm{PE}_{j,2i+1} &= \cos\left(\frac{j}{10000^{2i/D}}\right),
\end{align}
where $j$ is the index in the sequence and $i$ indexes the dimension $D$.
We discard, for the moment, the use of any positional encoding extracted from the wavelength position, but
we anticipate that it will be useful when lines observed with different instruments are used.
Finally, the output of the encoder is a sequence of latent vectors $z_i$ of dimension $D=256$ that represent the 
information extracted from the Stokes parameters, that will be utilized by the decoder to produce the model parameters.

The decoder is, again, a stack of
four transformer decoder blocks, each of which contains a masked multi-head self-attention layer, a multi-head
self-attention layer, and a feed-forward neural network. The masked self-attention layer allows the model to
learn the relationships between the input tokens to the decoder, only attending to the 
tokens before the one under consideration, so that the proper conditional probability distribution of
Eq. (\ref{eq:categorical}) is learned. The self-attention layer allows the model to
attend to the $z_i$ encoder latent vectors and learn how they affect the output sequence. The output of the decoder is a sequence
of categorical probability distributions over the tokens $\bar{e}_i$ in the codebook $\mathcal{C}$ for 
the model stratifications. The 
decoder produces the output sequence in an autoregressive way, predicting the next token in the sequence
given the previous tokens and the latent vectors $z_i$. In order to initiate the generation process, the decoder is
called with a \texttt{SOS} token, and the output sequence is generated one token at a time until the \texttt{EOS} 
token is produced. Positional encoding is also added to the input sequence of the decoder.

\begin{figure}
  \centering
  \sidecaption
  \includegraphics[width=\columnwidth]{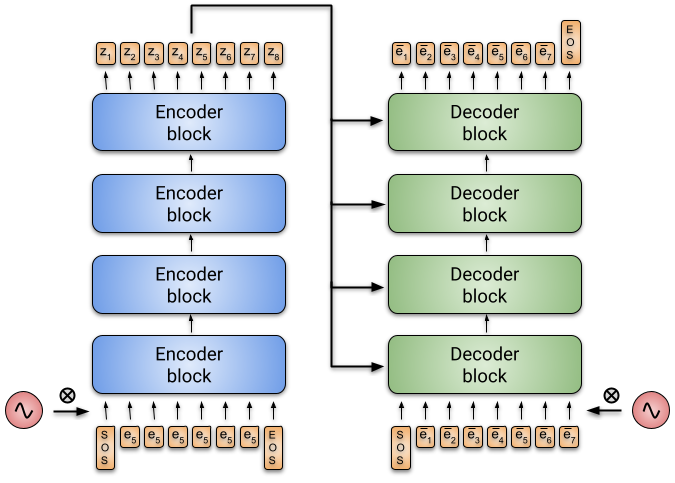}%
  \caption{Transformer encoder-decoder model. Both the encoder and decoder blocks are
  transformer layers..}
  \label{fig:transformer}
\end{figure}

\begin{figure*}
  \centering
  \sidecaption
  \includegraphics[width=0.9\textwidth]{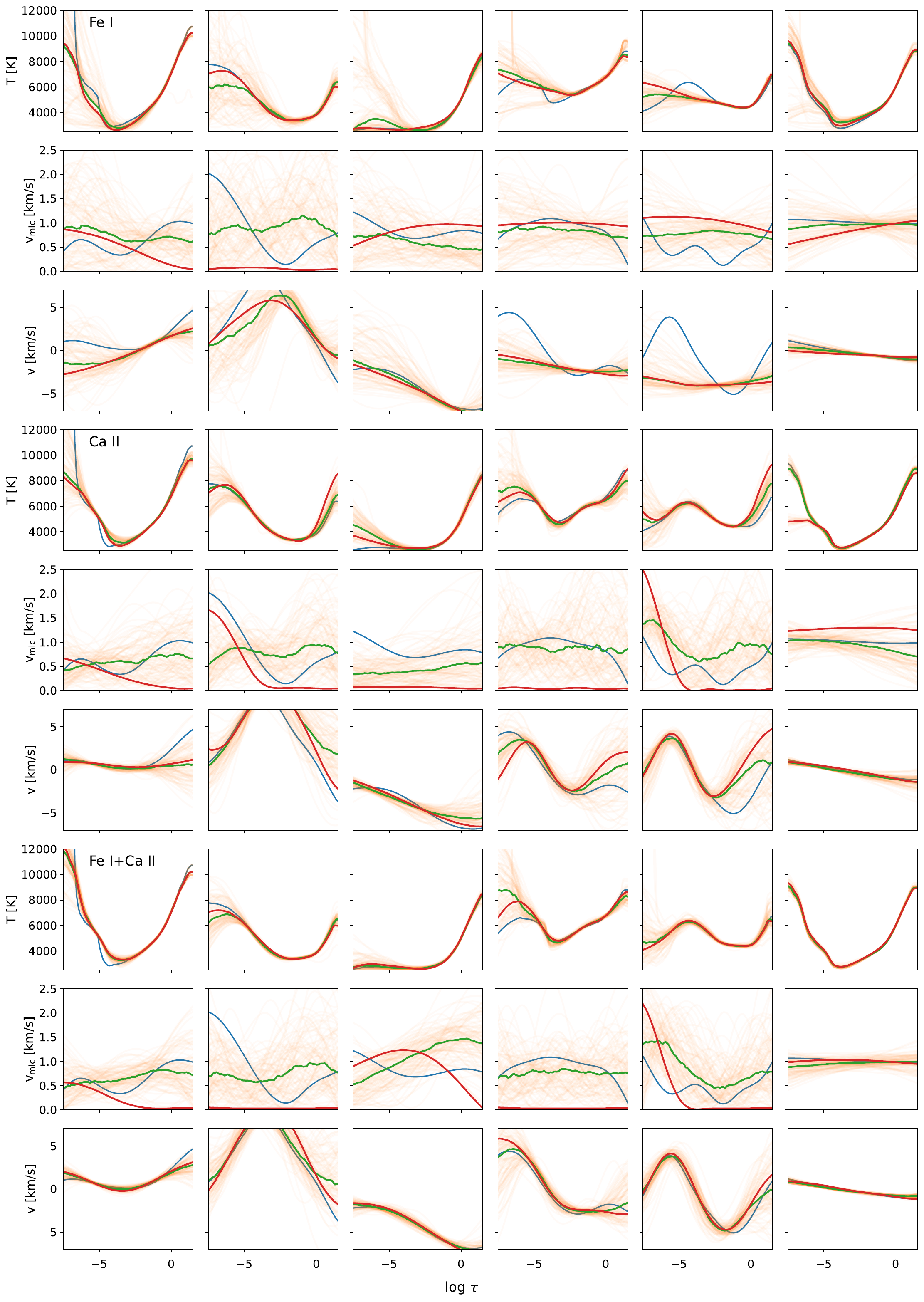}
  \caption{Samples from the generative model (orange curves) for the inversion of six Stokes profiles 
  extracted from the test set. The blue curve is the original model, the green curve is the median model
  and the red curve is the greedy decoded model. The upper three rows show the results when the Fe \textsc{i}
  spectral region is used, the middle three rows show the results when the Ca \textsc{ii} spectral region is used, and 
  the last three rows show the results when both spectral regions are used.}
  \label{fig:validation_inversion_thermo}
\end{figure*}

\begin{figure*}
  \centering
  \sidecaption
  \includegraphics[width=0.9\textwidth]{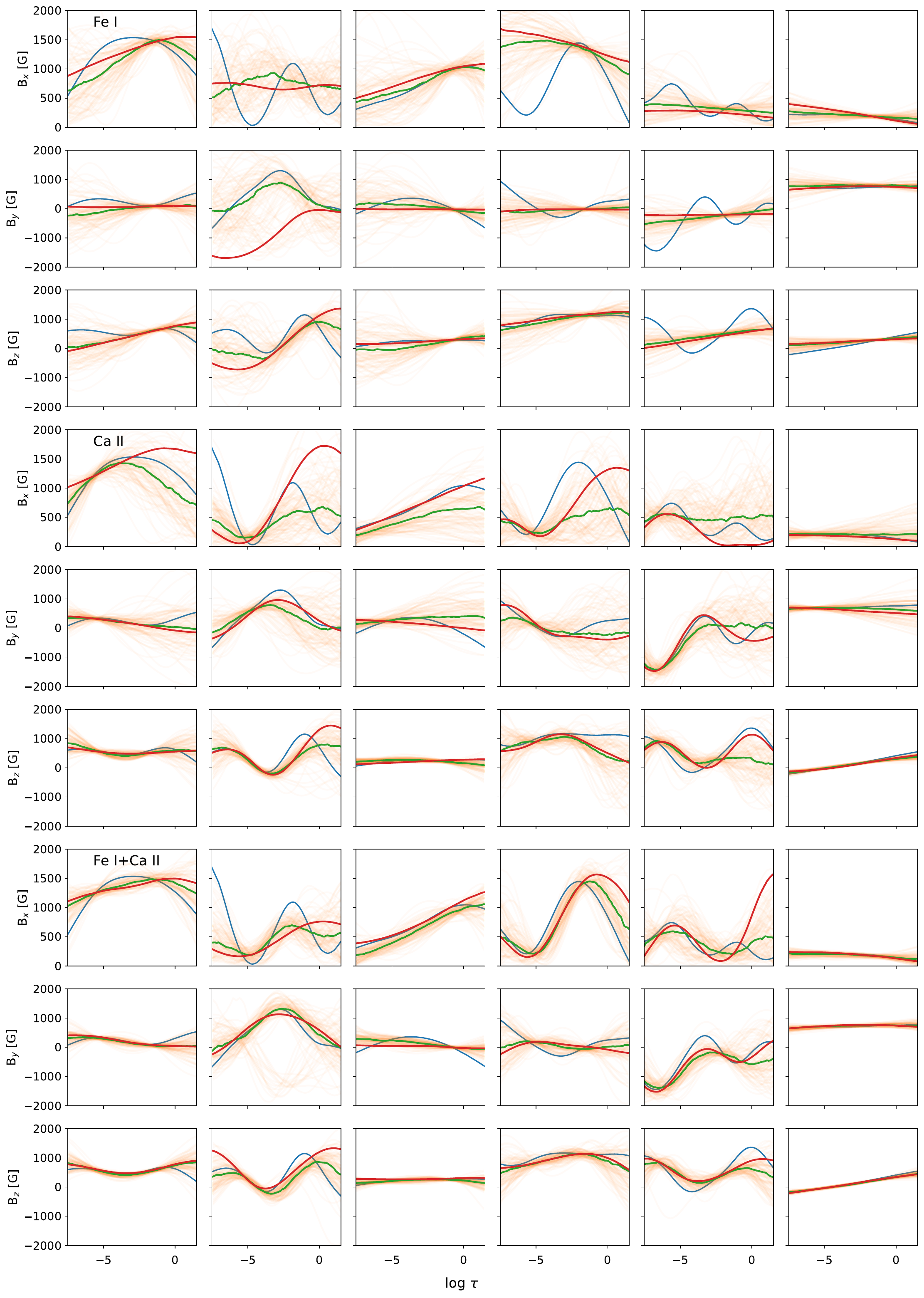}
  \caption{Same of Fig. \ref{fig:validation_inversion_thermo} but for the magnetic field components.}
  \label{fig:validation_inversion_magnetic}
\end{figure*}

\begin{figure*}
  \centering
  \sidecaption
  \includegraphics[width=\textwidth]{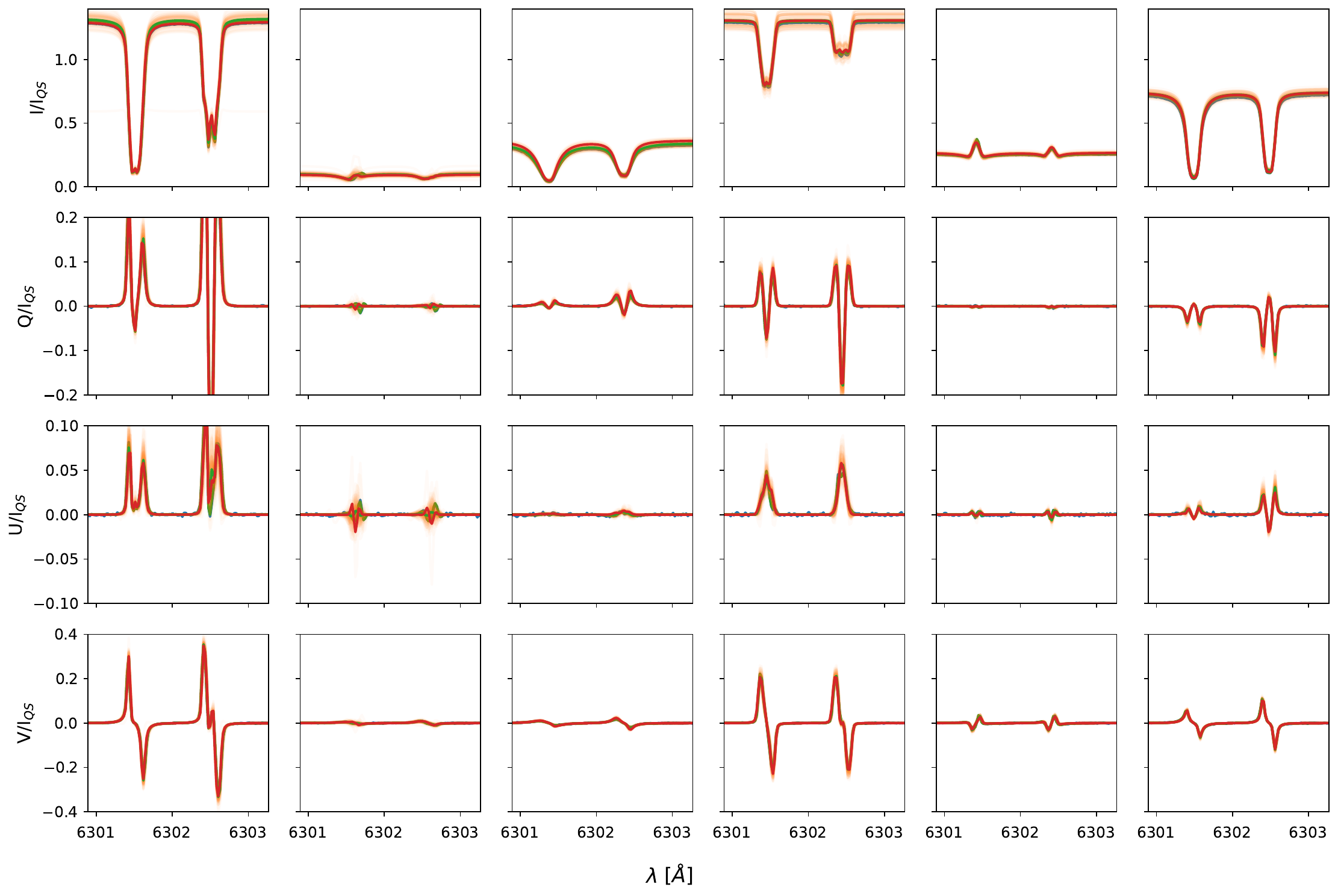}
  \caption{Synthesis of the Stokes profiles (orange curves) emerging from the models shown in Figs. \ref{fig:validation_inversion_thermo} 
  and \ref{fig:validation_inversion_magnetic}. The
  blue curves are the original Stokes profiles, the green and red curves are the median Stokes profiles and the greedy
  decoded Stokes profiles, respectively.}
  \label{fig:validation_inversion_stokes}
\end{figure*}

\begin{figure*}
  \centering
  \sidecaption
  \includegraphics[width=0.95\textwidth]{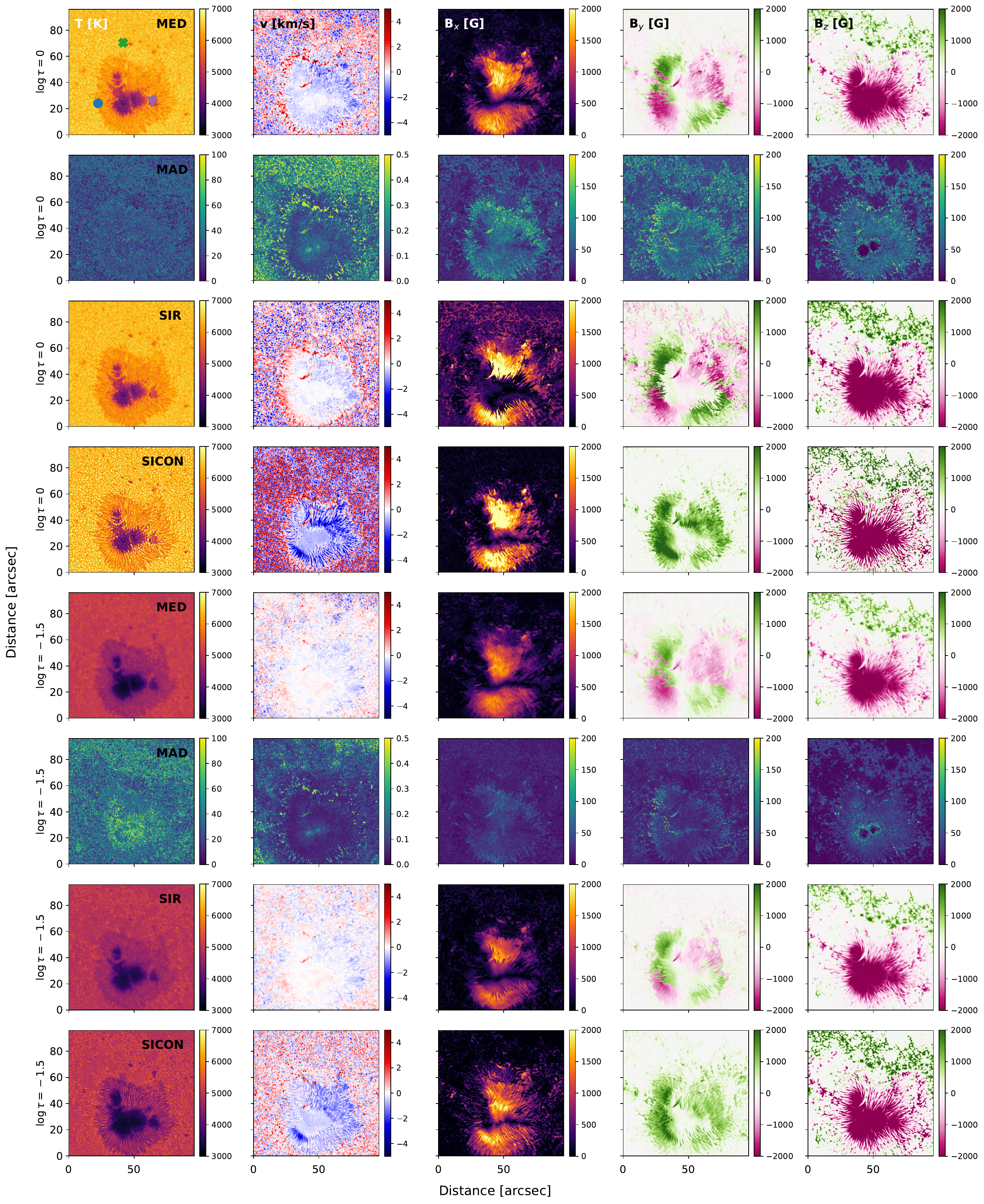}
  \caption{Inversion results for the AR10953.}
  \label{fig:hinode_images}
\end{figure*}

\begin{figure*}
  \centering
  \sidecaption
  \includegraphics[width=\textwidth]{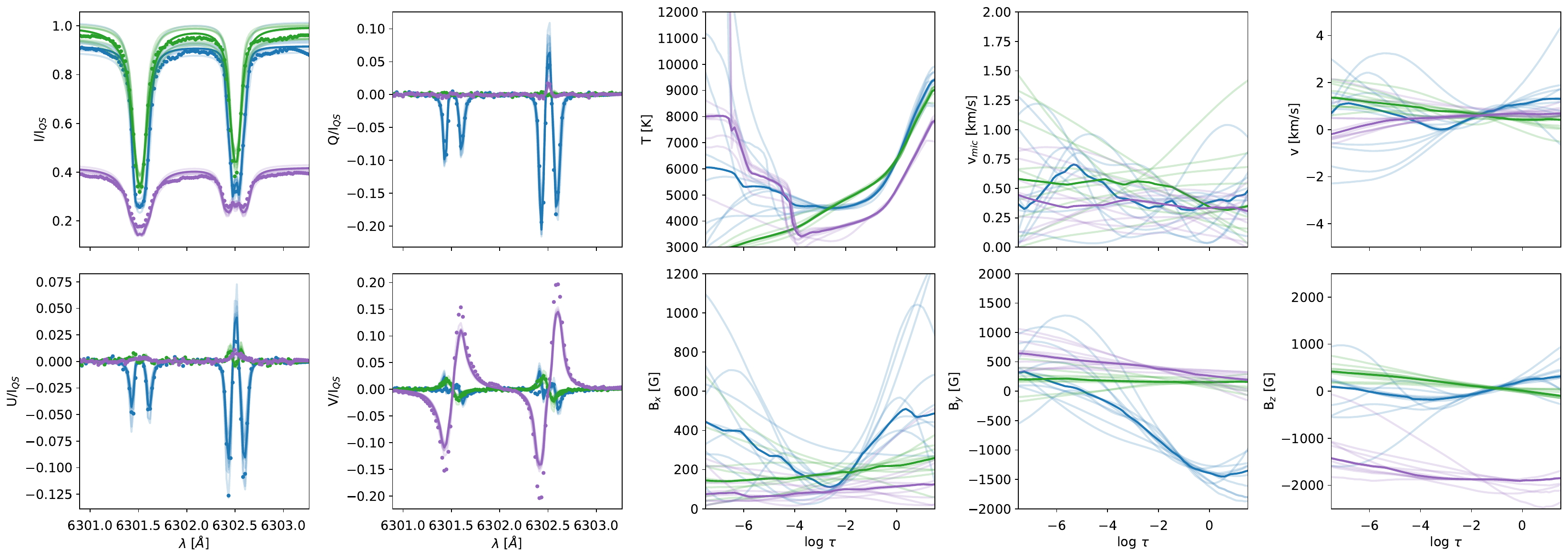}
  \caption{Synthesis of the Stokes profiles in the models inferred by \code\ in the three pixels
  marked in Fig. \ref{fig:hinode_images}. The dots indicate the observed Stokes profiles, the semi-transparent
  curves are the samples from the generative model, while the solid curves are the median of the samples. The colors
  correspond to the symbols in Fig. \ref{fig:hinode_images}.}
  \label{fig:hinode_sampling}
\end{figure*}

The transformer encoder-decoder architecture is shown in Fig. \ref{fig:transformer}. The transformer blocks 
are implemented using the \texttt{nn.TransformerEncoderLayer} and \texttt{nn.TransformerDecoderLayer} classes in \texttt{PyTorch}.
The transformer encoder-decoder is trained to minimize the cross-entropy loss between the output of 
the decoder and the target model parameters, which are also represented in the tokenized form. 
We use the Adam optimizer for the training, with a learning rate that is 
warmed-up from $3 \times 10^{-7}$ 
to $3 \times 10^{-4}$ in the first 50 epochs. It is then annaled using a cosine law until reaching $3 \times 10^{-5}$ after 1000 epochs.
We use a batch size of 512 and train the model in mixed precision, using half precision (\texttt{fp16}) and loss scaling, 
which takes advantage of the increased speed and reduced memory impact in modern GPUs, like the NVIDIA L40s and RTX4090
used in this work.

\section{Validation}
\label{sec:validation}

\subsection{Inference of model stratifications}
In order to see the capabilities of the trained model, we use the test set of 6000 models and their 
corresponding Stokes profiles. Although extracted from the same distribution as the training 
set, they have never been used during training. 
We explore two options to produce the output autoregressive sequence with the decoder. The 
first one is greedy decoding, in which the token with the highest probability
is selected at each step. This produces a single deterministic output model. The second one is sampling, in which every token 
is sampled from the estimated categorical distribution at each step. Since this token is used as input
to the decoder in the next step, the output is stochastic. When the decoder
is called many times, it produces an ensemble of model stratifications that can be used to estimate the uncertainty 
in the solution.

Figures \ref{fig:validation_inversion_thermo} and \ref{fig:validation_inversion_magnetic} show the 
results for the stratification of thermodynamical and magnetic quantities
inferred from six profiles examples from the test set, respectively. In both figures, the upper three rows display the results when
only the Fe \textsc{i} doublet at 630 nm is used, the middle three rows show the results when only 
the Ca \textsc{ii} line at 8542 \AA\ is used, and the last three rows show the results when both spectral regions 
are used. The blue curve is the original model stratification. The red curve
is the result from the greedy decoding. The orange curves are 100 samples from the generative model. These
models are represented with a low opacity, so that the zones with higher probability are more visible.
The green curve is the median of the samples. 

Focusing on the results of the Fe \textsc{i} doublet, we see that both the greedy solution and the median of the 
samples obtained from the distribution produce a good approximation to the original model in the line formation region for the 
doublet, which is located between $\log \tau_{500}=0$ and $\log \tau_{500}=-2.5$ \citep[see, e.g.,][]{2010A&A...518A...3O,2014A&A...572A..54B}. The
uncertainty in the solution is also much smaller in these regions than in the rest of the atmosphere. The 
contrary happens in the upper photosphere and chromosphere, where the uncertainties become larger 
because the Fe \textsc{i} doublet is barely sensitive to the specific physical conditions in these regions.
Despite this fact, the greedy and the median solution often overlap with the original model in the upper
regions of the atmosphere. We think that this is simply a consequence of the prior distribution 
inherent to the training data, which is based on the perturbation of semi-empirical models. These 
models already contain strong height correlations (e.g., there is a temperature rise in the chromosphere if 
the lower part of the atmosphere is similar to that of a quiet Sun model). If the construction of the training data
avoids using these semi-empirical models as a baseline, the resulting inferred models will surely be less 
informative in the upper layers of the atmosphere.

The results when the Ca \textsc{ii} line at 8542 \AA\ is used 
clearly demonstrate that the chromospheric line contains much more information about the
physical conditions in the upper layers of the atmosphere, so that the inferred models are much more constrained
in the chromosphere. We remind the reader that the core of the Ca \textsc{ii} line is sensitive to heights in
the range between $\log \tau_{500}\approx-1.5$ and $\log \tau_{500}\approx-4.0$ \citep[see, e.g.,][]{2016MNRAS.459.3363Q}.
Meanwhile, the uncertainties in the deeper layers of the atmosphere are 
larger than in the case of the Fe \textsc{i} doublet, a consequence of the fact that the Ca \textsc{ii} line is
very broad and the synthesized wavelength range does not reach continuum wavelengths. The inference in both spectral regions 
clearly demonstrate that the microturbulent velocity is not very well constrained, while the magnetic field 
vector seems to be better constrained with the chromospheric line.

Although not surprising, it is worth noting that the generative model has learned to produce model atmospheres
that are much more consistent with the original ones when the two spectral regions are used together. The 
inferred models look like a combination of the results obtained with the Fe \textsc{i} doublet and the Ca \textsc{ii} line
separately. The uncertainties are small in the deeper layers of the atmosphere thanks to the information provided 
by the Fe \textsc{i} doublet, while the chromospheric line provides a much better constraint in the upper layers of the atmosphere. 
The case of the temperature and LOS velocity stratifications is particularly interesting, as some of the inferred models
contain almost no uncertainty throughout the whole atmosphere (e.g., the first and last column). Other cases
remain uncertain in the upper layers of the atmosphere (e.g., fourth column). 

\subsection{Posterior check of synthetic Stokes profiles}
Bypassing the whole non-linear inversion of the radiative transfer equation has the undesired
consequence of not explicitly forcing the Stokes profiles to be fitted during the process, contrary to 
what happens with classical iterative inversion methods \citep[see, e.g.,][]{2019A&A...626A.102A}. This verification needs to be done
a posteriori, by synthesizing the Stokes profiles from the inferred model and comparing them with the original
Stokes profiles. If no good fitting is obtained, there is currently no way to improve the ML solution, except
arguably using the solution provided by the ML method as a first guess for a classical inversion code \citep{2021A&A...651A..31G}.
Other options can be based, for instance, on test-time computing (making the output of the ML model self-correct 
by using the Stokes profiles as input). Although they are very interesting and need to be explored in the future, 
they are outside the scope of this work.

Figure \ref{fig:validation_inversion_stokes} shows the results of the synthesis of the Fe \textsc{i} doublet on the inferred 
models using the Fe \textsc{i} observations in the test set shown in 
Figs. \ref{fig:validation_inversion_thermo} and \ref{fig:validation_inversion_magnetic}. The blue curves are the 
original Stokes profiles, the green and red curves are obtained in the median and greedy-decoded models, while 
the orange curves represent synthesis in the samples from the generative model. The synthesis is done using \texttt{Hazel2} under 
the assumption of LTE, in the same conditions of the generation of the training set. The similarity
between the original and the synthesized Stokes profiles is very good, reinforcing the fact that the
relevant part of the inferred models is well constrained by the model.

\section{Inversion of Hinode SOT/SP observations}
\label{sec:hinode}
The results of the validation are promising, so we apply the trained model to the inversion of Hinode SOT/SP observations.
We invert the observations of AR10953, which were obtained on 2007-05-01 almost at disk center. We use the 
scan starting at 16:30. The observations have been obtained in the fast scanning mode, in which the pixel 
size is 0.32 arcsec. The level-1 data is used without any modification, except for normalizing by the
intensity of the surrounding quiet Sun. Once the Stokes profiles are properly normalized, they are 
tokenized using the pre-trained VQ-VAE, passed to the encoder, and then
decoded by the decoder. The output sequence is then decoded using the VQ-VAE decoder to produce the model parameters.

We apply \code\ to the whole field of view (FoV), which contains 512$\times$512 pixels. Figure \ref{fig:hinode_images}
shows the most interesting subfield of the whole FoV. We display the inferred physical quantities at two
heights in the atmosphere, $\log \tau_{500}=0$, which is very close to the formation height
of the continuum (first four rows), and $\log \tau_{500}=-1.5$, which is closer to the formation height of 
the core of the lines (last four rows). The information obtained by our model at each optical depth surface is 
summarized with the median (MED) of the samples, together with the median absolute deviation (MAD) of the samples, which
is an estimator of the standard deviation that is robust to outliers. We use robust estimates because we 
have verified that some of the stratifications
produced by the model can be considered outliers. This is the equivalent of the so-called hallucinations that
are often found in generative models, which could eventually be reduced by using increasingly larger training sets.
We compare the results of the new inversion code with those obtained using the classical \texttt{SIR} inversion code (in our case, via 
the \texttt{Hazel2} inversion code), displayed in the third and seventh rows, and those obtained 
with the \texttt{SICON} inversion code \citep{2019A&A...626A.102A}, in the fourth and eighth rows. Note that the results
of \texttt{SICON} are crispier because the code automatically applies a deconvolution to the Stokes profiles, which is not
the case of \texttt{SIR} or the new inversion code. If one is interested in spatially deconvolved data, 
it is straightforward to use the approach of \cite{2013A&A...549L...4R} \citep[see also][]{2015A&A...579A...3Q}, by
first applying a regularized deconvolution to the Stokes profiles and then inverting them.

The resemblance between the inferred models and the \texttt{SIR} models is very good. Concerning the temperature, the
contrast of the plage surrounding the sunspot is slightly lower in our inversion than in the \texttt{SIR} inversion when
observed at $\log \tau_{500}=-1.5$. However, it is also true that the uncertainty is larger in these zones. On the
contrary, the estimation of the temperature close at $\log \tau_{500}=0$ is very certain. Likewise, 
the LOS velocity maps also correlate very well. In order to remove potential differences in the zero
point of the LOS velocity, we have subtracted the mean value of the LOS velocity in the FoV. Strong redshifts
are found in the limb-side penumbra (located on the left side of the images) in both inversions. Larger 
uncertainties are correlated with the strong redshifts. However, \texttt{SIR} also recovers redshifts in the
center-side penumbra (located on the right side of the images), which are not present in our inversion, even
after taking into account the uncertainties. The strong redshift of the most notable light bridge is common 
in the two inversions.

Concerning the magnetic field, strong similarities are found for all cartesian components. One obvious 
advantage of \code\ is the removal of the artifacts that are seen in the \texttt{SIR} inversion
in the $B_x$ and $B_y$ components, that are strongly correlated with the location of the plage. This
is a consequence of the well-known bias in the estimation of the transverse magnetic field from 
noisy Stokes $Q$ and $U$ profiles \citep{2012MNRAS.419..153M}, which is proportional to 
the broadening of the spectral line. This 
bias is also absent in the \texttt{SICON} inversion, in this case because of the exploitation of the
spatial correlation of the magnetic field vector. However, although the inferred $B_x$ component
is similar for the three codes, the $B_y$ component is very different with \texttt{SICON}. This points out
that the azimuths of the field provided by \code\ are more similar to those of \texttt{SIR} than those
of \texttt{SICON}. Concerning the longitudinal component of the 
magnetic field, the results are also very similar. The only difference is that the inferred 
$B_z$ component in \texttt{SIR} is slightly larger in the plage regions. Finally, since the observation is of good 
quality and the Stokes parameters have a good signal-to-noise ratio, the results
show good spatial coherence even though the inversions are carried out pixel-by-pixel.

\begin{figure*}
  \centering
  \sidecaption
  \includegraphics[width=\textwidth]{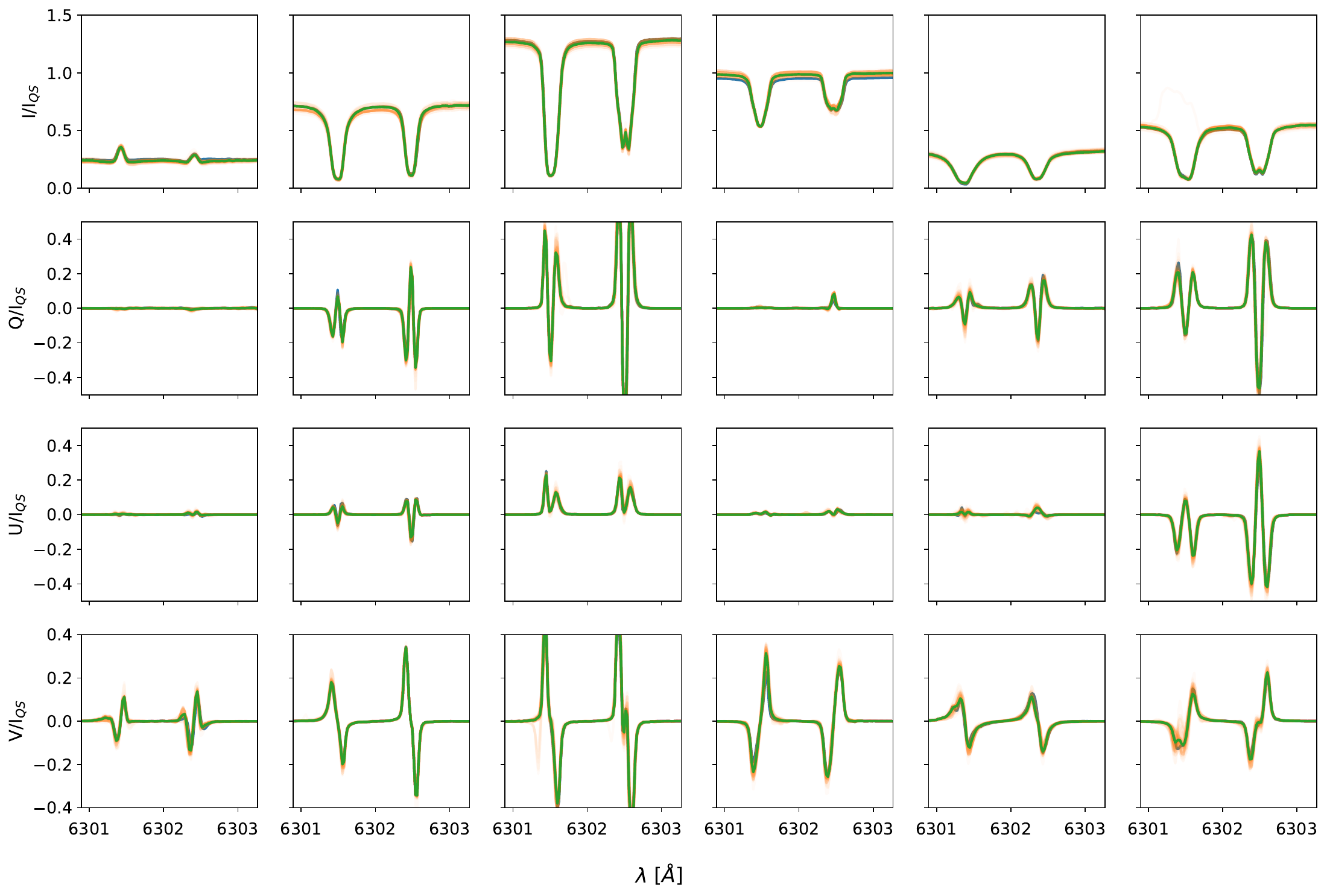}
  \caption{Samples from the generative model of the synthesis neural translation model.}
  \label{fig:validation_synthesis}
\end{figure*}

In order to further validate the results, we have selected three pixels in the FoV, marked with
blue, green and purple symbols in Fig. \ref{fig:hinode_images}, for which we show the results
in detail. The observed Stokes profiles in 
these pixels are shown with dots in the leftmost four panels of Fig. \ref{fig:hinode_sampling}. The semi-transparent
curves show the Stokes profiles synthesized in the models inferred with \code, while the median of the synthetic 
Stokes profiles is shown in solid color. The results display a very good agreement between the observed
and the synthesized Stokes profiles, even though this is a fully a posteriori check. The inferred
models (with 10 montecarlo realizations) are shown in the rightmost panels of Fig. \ref{fig:hinode_sampling}. Again,
the distributions have reduced uncertainty in the line formation region, while the uncertainty increases 
towards the upper layers of the atmosphere.

\section{Fast synthesis}
\label{sec:synthesis}
As commented before, given that the encoder-decoder architecture of Fig. \ref{fig:transformer} is a sequence-to-sequence model, 
it can also be used in reverse. In other words, the tokenized model stratifications can
be used as input, while the output of the model are the tokenized Stokes parameters. After passing
through the VQ-VAE decoder, they are transformed to Stokes parameters for the spectral region of
interest. This is, in summary, a generative model for the Stokes parameters, a stochastic synthesizer. Such a model 
can then be used for different purposes: fast synthesis of Stokes profiles for a large number of models,
markov chain Monte Carlo (MCMC) sampling of the posterior distribution for Bayesian inference, etc.
A summary of the results is displayed in Fig. \ref{fig:validation_synthesis}. The blue curves are the
original Stokes profiles in the model stratification of six models from the test set, the orange curves
are the samples from the generative model, while the green curves represents the median of the samples.
The green and blue curve overlap almost perfectly. We see, though, some large uncertainties (for instance, the
circular polarization profile of the last column). Our hypothesis is that these artifacts can be 
reduced by using a larger training set and/or usign a larger encoder-decoder model (e.g., more transformer blocks, larger
dimension of the latent vectors, etc.).

\section{Conclusions}
\label{sec:conclusions}
We have explored the use of autoregressive neural machine translation models to accelerate 
the inversion of Stokes profiles. The model is based on the use of a vector quantized variational autoencoder 
for tokenization of the Stokes parameters and the model stratifications. Stokes profiles and model stratifications
are then represented as sequences of tokens, like sentences in a natural language. The translation
is performed using a transformer encoder-decoder architecture under an autoregressive approach, which 
becomes a generative model given the stochastic nature of the decoder.

We have demonstrated that the trained model can be used to reliably infer the stratification of 
physical conditions in the solar atmosphere. The results are compatible with those obtained with
classical inversion codes, but the inference is much faster. The model does not 
depend on the complexity of the line formation process, so it can be used to quickly infer the thermodynamical
and magnetic stratification from lines whose formation is dominated by non-LTE or partial frequency redistribution 
effects. Additionally, given the generative
character of the model, it provides a way of sampling from the probability distribution of the model stratification,
allowing us to estimate the uncertainty in the solution. We have also demonstrated that the 
model can be used to carry out multi-line inversions by simply concatenating the tokenized
Stokes parameters of the different spectral lines. The model produces much better
constrained models when more spectral lines are added. This reveals that the model has learned to
exploit the information provided by the different spectral lines in a complementary way.
Finally, if the encoder-decoder model is used in reverse, it can also be trained
for the fast synthesis of Stokes profiles.

Even though the model is fast, transformer-based architectures are still computationally expensive
because of the quadratic scaling of the self-attention mechanism with the number of tokens in the sequence.
Given that current state-of-the-art natural language models are based on exactly the same architecture,
many techniques have been developed to reduce the computational cost of the model. For example, one can
use KV caching \citep{2022arXiv221105102P}, which allows the model to reuse the calculations performed for 
the attention when generating the next token, or linear attention mechanisms \citep[e.g.,][]{shen2019efficient}, 
which reduce the computational cost of the attention mechanisms.

The model presented in this work is of the encoder-decoder type, as usual in neural machine translation. 
However, recent studies have shown that decoder-only architectures can also be used for autoregressive translation 
\citep[e.g.,][]{2022arXiv221011807G,2023arXiv230404052F}. Using only a decoder could reduce the 
computational cost of the model since the encoder is not needed. The decoder-only architecture can be 
trained by concatenating the tokenized Stokes parameters and the model stratifications, and then
predicting the next token in the sequence given the previous tokens. The model can then be used
to produce the model stratifications when the Stokes parameters are provided as context. We will 
explore this architecture in future works.

Sequence-to-sequence models represent a very powerful tool for the inversion of Stokes profiles, given
the arbitrary character of the sequence. These models can seamlessly deal with the concatenation of
the tokenization of Stokes parameters of several spectral lines at arbitrary wavelength samplings, provided
that the relevant information is passed to the encoder. The main obstacle is the development of a dictionary
of tokens that can deal with arbitrary spectral lines at arbitrary wavelength samplings for 
arbitrary instruments. One can think of using a single VQ-VAE
tokenization for all spectral lines, modifying the encoder and decoder to deal with the arbitrary wavelength sampling.
Either 1D convolutional layers with appended wavelength information or transformer encoder and decoders could 
be used for this purpose. Another option is to use a different VQ-VAE tokenization for each spectral line, which would
allow the model to learn the relevant information for each spectral line independently, at the expense of
increasing the codebook size. We anticipate that tokenization coupled with generative models (either 
autoregressive or based on other ideas like diffusion models) will produce significant advances in the 
inversion of Stokes profiles in the near future, both for single pixel inversions and for
the inversion of large data cubes.

\begin{acknowledgements}
AAR acknowledges funding from the Agencia Estatal de Investigación del Ministerio de Ciencia, Innovación y 
Universidades (MCIU/AEI) under grant ``Polarimetric Inference of Magnetic Fields'' and the European Regional 
Development Fund (ERDF) with reference PID2022-136563NB-I00/10.13039/501100011033.
The Institute for Solar Physics is supported by a grant for research infrastructures of national importance from the Swedish Research Council (registration number 2021-00169). JdlCR gratefully acknowledges funding by the European Union through the European Research Council (ERC) under the Horizon Europe program (MAGHEAT, grant agreement 101088184).
We also acknowledge the contribution of the IAC High-Performance
Computing support team and hardware facilities to the results of this research.
Part of the computations were enabled by resources provided by the National Academic Infrastructure for Supercomputing in Sweden (NAISS), partially funded by the Swedish Research Council through grant agreement no. 2022-06725, at the PDC Center for High Performance Computing, KTH Royal Institute of Technology (project numbers NAISS 2024/1-14 and 2025/1-9)
This research has made use of NASA's Astrophysics Data System Bibliographic Services.
We acknowledge the community effort devoted to the development of the following open-source packages that were
used in this work: \texttt{numpy} \citep[\texttt{numpy.org},][]{numpy20}, 
\texttt{matplotlib} \citep[\texttt{matplotlib.org},][]{matplotlib}, \texttt{PyTorch} 
\citep[\texttt{pytorch.org},][]{pytorch19}, \texttt{scipy} \citep[\texttt{scipy.org},][]{2020SciPy-NMeth} and \texttt{scikit-learn} \citep[\texttt{scikit-learn.org},][]{scikit-learn}.
\end{acknowledgements}

%
%

\end{document}